\documentclass[twocolumn,english,aps,prd,reprint,floatfix,notitlepage,footinbib,preprintnumbers,superscriptaddress]{revtex4-1}
\pdfoutput=1
\usepackage{lmodern}

\usepackage[T1]{fontenc}
\usepackage[latin9]{inputenc}
\usepackage{geometry}
\usepackage{subfigure,lmodern, amsmath,amssymb, graphicx, pifont, adjustbox, bm, xcolor}
\usepackage{amsfonts}
\usepackage{geometry}
\usepackage{comment}
\usepackage{mathtools}
\usepackage{float}
\usepackage{slashed}
\usepackage{ragged2e}
\usepackage{array}
\usepackage{hhline}

\makeatletter\g@addto@macro\bfseries{\boldmath}\makeatother

\newcommand{\SigmaScaled}{\scalebox{1.85}{\raisebox{-0.65mm}{$\Sigma$}}}

\usepackage{stackengine}
\usepackage{esint}
\usepackage[unicode=true,pdfusetitle,
 bookmarks=true,bookmarksnumbered=false,bookmarksopen=false,
 breaklinks=false,pdfborder={0 0 1},backref=false,colorlinks=true]
 {hyperref}
\hypersetup{
 pdfauthor={Clifford Cheung, Aaron Hillman, and Grant N. Remmen},
 citecolor=black,linkcolor=black,urlcolor=black}

\newcommand{\appendixref}[1]{\hyperref[#1]{appendix~\ref{#1}}}
\def\equationautorefname~#1\null{eq.\,(#1)\null}
\usepackage{breakurl}
\usepackage{breakurl}
\usepackage[hang,flushmargin]{footmisc} 
\allowdisplaybreaks
\makeatletter

\usepackage{etoolbox}
\apptocmd{\thebibliography}{\justifying\setlength{\leftskip}{7.4mm}}{}{} 
 
 \usepackage{relsize}
\usepackage{babel}

\makeatletter
\def\simgt{\mathrel{\lower2.5pt\vbox{\lineskip=0pt\baselineskip=0pt
           \hbox{$>$}\hbox{$\sim$}}}}
\def\simlt{\mathrel{\lower2.5pt\vbox{\lineskip=0pt\baselineskip=0pt
           \hbox{$<$}\hbox{$\sim$}}}}
\makeatother

\usepackage{changepage}

\newcommand{\be}{\begin{equation}}
\newcommand{\ee}{\end{equation}}
\newcommand{\bea}{\begin{eqnarray}}
\newcommand{\eea}{\end{eqnarray}}
\newcommand{\Fig}[1]{Fig.~\ref{#1}}

\newcommand{\Eq}[1]{Eq.~(\ref{#1})}

\newcommand{\App}[1]{App.~\ref{#1}}

\newcommand{\eq}[2]{\be\begin{aligned}#1 \label{#2}\end{aligned}\ee}

\newcommand{\mysec}[1]{\noindent {\bf #1.}---}
\newcommand{\mysubsec}[1]{\noindent {\it #1.}---}

\newcolumntype{P}[1]{>{\centering\arraybackslash}p{#1}}

	\definecolor{dartmouthgreen}{rgb}{0.05, 0.5, 0.06}

\newcommand{\mm}{\mu}
\newcommand{\xx}{\xi}
\newcommand{\DD}{w}

\begin{document}

\preprint{CALT-TH 2024-022}

\title{Bootstrap Principle for the Spectrum and Scattering of Strings}

\author{Clifford Cheung}
\affiliation{Walter Burke Institute for Theoretical Physics, California Institute of Technology, Pasadena, CA 91125}
\author{Aaron Hillman}
\affiliation{Walter Burke Institute for Theoretical Physics, California Institute of Technology, Pasadena, CA 91125}
\author{Grant N. Remmen}
\affiliation{Center for Cosmology and Particle Physics, Department of Physics, New York University, New York, NY 10003}
    
\begin{abstract}
\noindent 
We show that the Veneziano amplitude of string theory is the unique solution to an analytically solvable bootstrap problem. Uniqueness follows from two assumptions: faster than power-law falloff  in high-energy scattering and the existence of some infinite sequence in momentum transfer at which higher-spin exchanges cancel. The string amplitude---including the mass spectrum---is an output of this bootstrap.  If the amplitude merely vanishes at high energies, the solution is a three-parameter family containing the Veneziano, Coon, and hypergeometric amplitudes, and more. 

\end{abstract}
\maketitle

\preprint{}

\maketitle

\mysec{Introduction}String theory is a rare working prototype for how to consistently reconcile quantum mechanics and gravity at arbitrarily short distances.  To do so, it asserts that the most basic building blocks of nature are vibrating strings rather than point particles.  But is this assumption in any sense unique or inevitable?

Naively, the only way to assess the uniqueness of string theory is to compare and contrast it against its alternatives, but these are few and far between.   A more systematic path is to apply the bootstrap principle, which aims to sculpt the observables of a theory directly from a set of conservative, bottom-up assumptions.  A particularly fruitful version of this idea is the scattering amplitudes bootstrap, which utilizes basic principles like Lorentz invariance, locality, and unitarity to construct the space of all possible scattering amplitudes, which in turn dictates the space of all possible theories \cite{Elvang:2015rqa,Dixon:1996wi, Cheung:2017pzi,Travaglini:2022uwo}.  In this approach, theories like quantum chromodynamics and general relativity are not derived from textbook gauge invariance and general covariance, but rather as the unique theories of the interactions of massless spin one and spin two particles that are Lorentz invariant, local, and unitary.
\medskip

\mysubsec{Assumptions}What are the bottom-up assumptions that fix the scattering amplitudes of string theory?  That is, without any preconceived notions about the string, like its detailed spectrum or dynamics, how would we arrive at those elements from a systematic bootstrap?   

We can look for hints to this question within the string amplitudes themselves, which famously exhibit a number of extraordinary properties.  Chief among these is their high-energy behavior, which does not diverge with center-of-mass energy but rather attenuates, and at a rate that is impossible in a quantum theory of point particles.  This falloff is encoded in the Regge limit, 
\eq{
\lim_{|s| \rightarrow\infty} A(s,t)  \sim s^{J(t)},
}{}
where $J(t)$ is the Regge trajectory.
When there exists some range of $t$ for which $J(t)<0$, as is the case for tree-level string amplitudes, an unsubtracted dispersion relation then implies that~\footnote{Throughout this paper we consider planar, color-ordered amplitudes, which arise in gauge theories or open string theory.},
\eq{
A(s,t) &= \oint\limits_{s'=s} \frac{{\rm d}s'}{2\pi i} \frac{A(s',t)}{s'-s}  = \int_{-\infty}^\infty \frac{{\rm d}s'}{\pi} \frac{{\rm Im}\,A(s',t)}{s'-s}.
}{eq:disp}
A priori, the amplitude can have discontinuities in both the $s$ and $t$ channels.  Nevertheless, it has been recast as a sum over solely $s$ channel contributions.  This property of {\it dual resonance}---that the amplitude can be expressed either as a sum over $s$ or $t$ channel exchanges---is implied by vanishing Regge behavior~\cite{Cheung:2023adk, Caron-Huot:2016icg, Cheung:2023uwn, Nakanishi:1970xz, Matsuda:1969cso}.  

Throughout this work, we will posit a weakly coupled tree-level amplitude, so the right-hand side of the dispersion relation is a sum over simple poles,
\be 
A(s,t) = \sum_{n=0}^\infty \frac{R(n,t)}{\mm(n)-s},\label{amp_dual_res}
\ee
where the sequence $\mm(n)$ parameterizes an a priori arbitrary spectrum indexed by a level number $n$, which runs over the nonnegative integers.    

By definition, the pole in Eq.~\eqref{amp_dual_res} at level $n$ comprises a set of exactly degenerate exchanged states of different spins.
The degree of each residue $R(n,t)$ with respect to $t$ dictates the maximal spin exchanged at level $n$. 
If $R(n,t)$ is nonpolynomial, then the exchanged state is of finite energy but {\it infinite} spin.  Any such mode has infinite extent, thus running afoul of locality~\cite{Huang:2022mdb,Bern:2021ppb}.  Notably, unphysical amplitudes of this type---for example, $A \propto 1/(s-m^2)(t-m^2)(u-m^2)$---appear as ``extremal'' amplitudes that are formally unitary but are nonlocal and frequently contaminate numerical bootstrap analyses~\cite{Caron-Huot:2020cmc, Caron-Huot:2021rmr, Bern:2021ppb}.  In order to prune away these pathological amplitudes, we impose the additional assumption of locality, so any exchanged state has finite spin and $R(n,t)$ is a polynomial.  
Crossing symmetry then implies that \Eq{amp_dual_res} must have poles in $s$ at arbitrarily high level $n$ in order to account for expansion of the poles in the $t$ channel. 
Assuming for simplicity that there are no distinct resonances with the same maximal spin, then without further loss of generality we can define the ordering of $n$ such that $n$ is the degree of $R(n,t)$, so $\mm(n)$ is a nonrepeating sequence and the sum in \Eq{amp_dual_res} runs over spin. 

The amplitudes of confining gauge theories and fundamental strings both have vanishing Regge behavior, but it is the strength of this falloff that distinguishes them.   For confining gauge theories, $J(t)$ asymptotes to a constant for negative $t$, corresponding to the impossibility of arbitrarily soft high-energy behavior in quantum field theory \cite{Martin, Buoninfante:2023dyd}.   
For fundamental strings, however, $J(t)$ decreases unboundedly for negative $t$.  In the parlance of Ref.~\cite{Haring:2023zwu}, such amplitudes are ``superpolynomially soft,'' which means that for any positive integer $k$ there exists some range of $t$ for which $A(s,t)$ is bounded by $s^{-k}$.  This implies that there exist some $t_k$ for which  $A(s,t_k)\sim s^{-k}$ in the Regge limit~\footnote{If the sequence $t_k$ is unboundedly negative for large $k$ then $t=t_k$ interpolates to the regime of high-energy fixed-angle scattering.  In this case, superpolynomial softness coincides with exponentially soft fixed-angle scattering, which is another calling card of string amplitudes.  A loophole to this logic is if $t_k$ approaches an accumulation point, which actually occurs for the Coon amplitude when $q>1$.}.  This scaling implies that the Regge trajectory $J(t)$ crosses all negative integers, so $J(t_k) = - k$.
Since $A(s,t_k)  \sim s^{-k}$, it is then natural to wonder if the amplitude is actually a {\it rational} function of $s$ at these points.  Indeed, this is precisely the case for the Veneziano amplitude~\footnote{The Veneziano amplitude referred to here and throughout describes the scattering of external states that are massless colored scalars.  This amplitude is obtained by a trivial affine shift of $s$ and $t$ in the bosonic string amplitude or, alternatively, by stripping off the polarization data from superstring scattering, which is also known as the $Z$-theory amplitude~\cite{Carrasco:2016ldy}.},\vspace{-1mm}
\be \vspace{-1mm}
A_V(s,t) = \frac{\Gamma(-s)\Gamma(-t)}{\Gamma(-s-t)},\label{eq:AV} 
\ee
which degenerates to a rational polynomial of degree $-k$ precisely when $t=-k$.
In particular, at these special values of the momentum transfer, the Veneziano amplitude exhibits a {\it finite} number of poles in the $s$ channel,\vspace{-1mm}
\be  \vspace{-1mm}
A_V(s,-k) = \sum_{n=0}^{k-1} \frac{(1-k)_n}{n!} \frac{1}{n-s},
\ee
where we have defined the Pochhammer symbol $(a)_n=  \prod_{k=0}^{n-1}(a\,{+}\,k) = \Gamma(a\,{+}\,n)/\Gamma(a)$.  Setting $t=-k$, the partial waves at each level $n\geq k$ destructively interfere, leaving no sign of those exchanged states in the amplitude.  We dub this phenomenon ``level truncation,'' where all terms at level $n\geq k$ in the dual resonant sum in Eq.~\eqref{amp_dual_res} vanish at some sequence of values of $t$ that we call $\xi(k)$,  leaving a finite sum over poles.
By definition, an amplitude that satisfies level truncation will be a rational polynomial at $t=\xi(k)$, with Regge behavior going at most like $s^{-1}$; while the tree-level string itself has softer Regge scaling $\sim s^{-k}$, we will not enforce this improvement for now.

\medskip

\mysubsec{Approach and Results}In this paper, we take this curious property of the Veneziano amplitude and elevate it the status of an {\it input assumption} of a bootstrap calculation.  Concretely, we present an analytic bootstrap for tree-level, crossing symmetric amplitudes $A(s,t)$ describing a tower of finite-spin modes whose spectrum is an unspecified sequence $\mm(n)$.  At the heart of our construction lie two nontrivial assumptions.  The first is that the amplitude has vanishing Regge behavior, so it can be expressed as a sum over poles in just the $s$ channel, as in \Eq{amp_dual_res}.  The second is level truncation, which states that this sum only receives contributions from a finite set of levels with $n<k$ when $t=\xx(k)$, where $\xx(k)$ is yet another unspecified sequence.

This bootstrap problem can be solved analytically and exactly.   Our construction imposes nonlinear constraints on the sequences $\xx(n)$ and $c(n)$ and on the spectrum $\mm(n)$.  Consequently, unlike all prior analyses in this subject, the spectrum $\mm(n)$ is  an {\it output} of our bootstrap rather than an input.   The general solution of this bootstrap problem is a three-parameter $(q,r,\DD)$ family of amplitudes.  The spectrum is  
\eq{
\mm(n) =[n]_q = \frac{1-q^n}{1-q},
}{q_spectrum}
 which is the $q$-integer spectrum of the Coon amplitude~\cite{Coon:1969yw}.  As we will find, $q\geq 1$ is mandated by convergence of the dual resonant sum, but unitarity is violated for $q>1$.  Hence, we are left with $q=1$, corresponding to the linear spectrum of the string.
Meanwhile, $r$ describes the parameter of the hypergeometric amplitudes proposed in Ref.~\cite{Cheung:2023adk}, and the remaining parameter $\DD$ characterizes a new class of amplitudes.     By construction, all of these amplitudes have vanishing Regge behavior but only for $r=\DD=0$ is the amplitude superpolynomially soft.  As a result, the unique object satisfying this last condition is precisely the Veneziano amplitude.
\medskip\vspace{0.8mm}

\mysec{Analytic Bootstrap}Vanishing Regge behavior implies that the amplitude admits the dual resonant representation in \Eq{amp_dual_res}.  We now show how level truncation constrains the residue and impose crossing symmetry.
\medskip

\mysubsec{Residue Structure}At $t=\xx(1)$, the amplitude only receives contributions from level $n=0$.  Consequently, all the other residues must vanish and are hence proportional to $t-\xx(1)$.  On the other hand, at $t=\xx(2)$ the only exchanges are from levels $n=0,1$, so the remaining residues are proportional to $t-\xx(2)$, and so forth.  Iterating this logic, we learn that the residues are
\eq{
R(n,t) = c(n) \prod_{k=1}^n (t-\xx(k)),
}{CRZ}
where $c(n)$ is an unspecified sequence. 

Together, Eqs.~\eqref{amp_dual_res} and \eqref{CRZ} describe an amplitude parameterized by three infinite sequences: the spectrum $\mm(n)$, the residue normalization $c(n)$, and the loci of level truncation $\xx(k)$.  The resulting amplitude has the same structure of coincident residue zeros as the Veneziano amplitude, but with the positions of all poles and zeroes in $s$ and $t$ transformed by the schematic replacement,
\eq{\
\textrm{integer} \;\;\;\longrightarrow\;\;\; \textrm{arbitrary sequence}.
}{schematic_replacement}
Said another way, our bootstrap automatically sculpts out a putative space of scattering amplitudes whose poles and residue zeros have the same overall structure as Veneziano, but at a priori different locations. 

Without loss of generality, let us consider all amplitudes with fixed overall normalization and modulo arbitrary transformations of $s$ and $t$ under shifts and rescaling.  We can eliminate these degrees of freedom by fixing $\mm(0)=0$, $\mm(1)=1$, and $c(0)=1$, which will reduce the complexity of our formulas.

A common obstacle in attempting to bootstrap string amplitudes from first principles is that most starting assumptions allow for the possibility of {\it sums of} Veneziano satellites \cite{Gross:1969db, Haring:2023zwu}, which are simply Veneziano amplitudes evaluated at shifted or rescaled values of $s$ and $t$.  Hence, one typically anticipates an infinite space of viable solutions to any given bootstrap.  However, in our case \Eq{CRZ} shows that the residues that are implied by our assumptions are very constrained.  A sum of Veneziano satellites will in general fail the condition required by \Eq{CRZ}.  This is a victory because it means that the solution space of our bootstrap problem will not be diluted by these infinite classes of amplitudes.
\medskip

\mysubsec{Crossing Symmetry}The requirement of crossing symmetry has yet to be imposed.   In general this is a formidable but in some cases achievable task~\cite{Eckner:2024ggx} because the amplitude is an infinite sum  of terms, as described in \Eq{amp_dual_res}.  Hence, crossing symmetry places an infinite number of constraints on an infinite number of parameters.
However, due to level truncation we know that the amplitude reduces to a finite collection of terms at certain kinematic points, in which case the conditions of crossing symmetry are similarly finite.

Consider the amplitude at $s=\xx(i)$ and $t=\xx(j)$ for some positive integers $i$ and $j$,
\eq{
A(\xx(i),\xx(j))=\sum_{n=0}^{j-1} \frac{c(n)\prod_{k=1}^n (\xx(j)-\xx(k))}{\mm(n)-\xx(i)}.
}{}
Imposing crossing symmetry, and assuming that $\xx(i)$ and $\xx(j)$ have real parts in the regime of convergence of the dual resonant sum, we have
\eq{
A(\xx(i),\xx(j))=A(\xx(j),\xx(i)),
}{eq:crossing}
mandating nonlinear constraints on the sequences $c(n)$, $\xx(n)$, and $\mm(n)$.
While these constraints are complicated, they only involve a finite number of sequence variables, and it turns out that we can solve them analytically.    

Without prior knowledge of the existence of string amplitudes, it would be surprising that these equations admit any solutions at all.  In particular, the constraint of crossing symmetry in \Eq{eq:crossing} at large $i$ and $j$ involves a maximal level $n_{\rm max} \sim \max(i,j)$.   The number of unknowns in the sequences $c(n)$, $\xx(n)$, and $\mm(n)$ up to this maximal level scales as $3n_{\rm max}$, while the number of equations in  \Eq{eq:crossing} scales as $n_{\rm max}^2$.   Hence, this system of nonlinear equations is exceedingly overconstrained.

While \Eq{eq:crossing} is a system of nonlinear equations, we can actually solve them sequentially as linear equations.
First, we determine the normalization sequence $c(n)$ by solving  \Eq{eq:crossing} in the case of $i\;{=}\;1$ and $j\;{\geq}\;2$, yielding \vspace{-3mm}
\be 
c(n)=(-1)^{n}\left(\xx(1)-\mm(n)\right)\prod_{k=1}^{n+1}(\xx(k))^{-1}.\vspace{-3mm}
\ee
Second, \Eq{eq:crossing} for $i\,{=}\,2$ and $j\,{\geq}\,3$ uniquely fixes the spectrum $\mu(n)$, whose solutions are defined by the recursion relation, 
\be 
\mu(n)=[\mu(n-1)(\xx(n+1)-1)+\xx(2)]/\xx(n+1).
\ee
 Last, \Eq{eq:crossing} for $i\;{=}\;3$ and $j\;{\geq}\; 4$ determines the loci of level truncation, which are recursively defined by 
\be 
\xi(n) = \xi(3) (\xi(n-1)-1)/(\xi(2)-1).
\ee
Note that the above manipulations assume that the sequence $\xi(n)$ is nonrepeating, which is equivalent to requiring that under level truncation the dual resonant sum at $t=\xx(k)$ receives contributions {\it at and only at} each pole $n\;{<}\;k$.
The resulting space of solutions depends solely on three remaining variables $\xx(1)$, $\xx(2)$, and $\xx(3)$, which we remap to new parameters $(q,r,w)$ via 
\be 
\begin{aligned}
\xx(1)&=[-1-r-\DD]_q,\;\;\xx(2)=[-2-r]_q,\\
\xx(3)&=[-3-r]_q.
\end{aligned}
\ee
The sequences are then
\eq{
\mm(n) &= [n]_q \\
\xx(n) &= [-n-r-\delta_{1n}\DD]_q\\
c(n) &= \frac{(1-q)^n q^{n(n+3+2r)/2}[1+n+r+\DD]_q}{(q^{2+r};q)_n [1+r+\DD]_q} ,
}{solutions}
where $(a;q)_n = \prod_{k=0}^{n-1}(1-aq^k)$ is the $q$-Pochhammer symbol.  Thus, we arrive at a three-parameter $(q,r,w)$ space of amplitudes.  Note that \Eq{eq:crossing} is necessary but not sufficient for crossing symmetry, and one must still check that the amplitudes corresponding to \Eq{solutions} are crossing symmetric at generic $s$ and $t$ for consistency.
\medskip

\mysec{Solution Space}The solutions in \Eq{solutions} encode the full space of scattering amplitudes that satisfy our bootstrap.  Let us now present explicit expressions for these amplitudes and study their high-energy behavior.
\medskip

\mysubsec{Amplitudes}For notational ease, let us first define convenient kinematic variables $\sigma = \log[1-(1-q)s]/\log q$ and
$\tau =\log[1-(1-q)t]/\log q$, which are constructed so that $[\sigma]_q = s$ and $[\tau]_q = t$.  Plugging \Eq{solutions}  into \Eq{CRZ}, we have the residues
\eq{\hspace{-1mm}
\! R(n,t) \,{=}\, \frac{q^{n}(q^{1{+}r{+}\tau};q)_{n}[1{+}r{+}\tau{+}\DD]_{q}[1{+}n{+}r{+}\DD]_{q}}{q^{\DD}(q^{2{+}r};q)_{n}[1{+}r{+}\tau]_{q} [1{+}r{+}\DD]_{q}} ,\hspace{-1mm}
}{R_sol}
for $n>0$, while the $n=0$ residue is $R(0,t)=1$.

We then insert the residue in \Eq{R_sol} into the dual resonant form of the amplitude in \Eq{amp_dual_res}. From the definition of the $q$-deformed generalized hypergeometric functions~\footnote{The $q$-deformed (basic) generalized hypergeometric functions are defined by the sum
\begin{equation*}
\;\qquad{}\,_{n+1}\phi_{n}\left[\begin{smallmatrix}
a_1,\ldots,a_{n+1}\\
b_1,\ldots,b_n
\end{smallmatrix};q;z\right]   = {\SigmaScaled}_{k=0}^\infty \tfrac{(a_1;q)_k \cdots (a_{n+1};q)_k}{(b_1;q)_k \cdots (b_n;q)_k}\tfrac{z^k}{(q;q)_k},
\end{equation*} 
and $\Gamma_q(z)=(1-q)^{1-z}(q;q)_\infty/(q^z;q)_\infty$ is the $q$-gamma function.} and their various identities, we obtain the amplitude in manifestly crossing symmetric form,
\begin{widetext}
\be\vspace{-2mm}
\begin{aligned}
A(s,t)= &\, \frac{[1+r]_{q}}{[1+r+\DD]_{q}}q^{\sigma\tau-\DD}\frac{\Gamma_{q}(-\sigma)\Gamma_{q}(-\tau)}{\Gamma_{q}(-\sigma-\tau)}\,_{3}\phi_{2}\left[\begin{array}{c}
q^{-\sigma},\,q^{-\tau},\,q^{r}\\
q^{-\sigma-\tau},\,q^{1+r}
\end{array};q;q\right]
+q^{-\DD}\frac{[\DD]_{q}\left([1+r+\DD]_{q}+[1+r]_{q}\right)}{[1+r+\DD]_{q}st}
\\ & -q^{\sigma\tau-\DD}\frac{[\DD]_{q}\left([1+r+\DD]_{q}+[1+r]_{q}-[-\sigma]_{q}-[-\tau]_{q}\right)}{[1+r+\DD]_{q}}\frac{\Gamma_{q}(-\sigma)\Gamma_{q}(-\tau)}{\Gamma_{q}(1-\sigma-\tau)}\,_{3}\phi_{2}\left[\begin{array}{c}
q^{-\sigma},\,q^{-\tau},\,q^{1+r}\\
q^{1-\sigma-\tau},\,q^{2+r}
\end{array};q;q\right].
\end{aligned}\label{eq:A}
\ee
\end{widetext}\vspace{-3mm}
Independently of $r$ and $\DD$, \Eq{eq:A} scales as $A\sim q^{\sigma\tau}$ in the Regge limit, which vanishes when ${\rm Re}(\tau)<0$~\footnote{A remarkable feature of this amplitude is the appearance of the $q^{\sigma\tau}$ prefactor.  This very same quantity was introduced as a nonunique~\cite{Jepsen:2023sia} fudge factor sitting in front of the amplitude in Ref.~\cite{Coon:1972qz} in order to cancel the nonpolynomiality of the residues of the original Coon amplitude~\cite{Coon:1969yw} for $q<1$.  Surprisingly, we find that this same prefactor arises {\it automatically} from the meromorphic dual resonant sum for the  amplitude for $q>1$, and need not be included by hand. Meanwhile, in the high-energy fixed-angle limit defined by $|s|,|t|\rightarrow \infty$ at fixed $t/s$,  we find that $\log A \sim \log s \log t/\log q$ for  $q > 1$.   This is the distinctive double-log scaling characteristic of the Coon amplitude~\cite{Caron-Huot:2016icg,Cheung:2022mkw}, but which we have found here to be independent of the new parameters $r$ and $\DD$.  Notably, such behavior for $q<1$ can arise in consistent ultraviolet theories, such as the anti-de Sitter/D-brane construction of Ref.~\cite{Maldacena:2022ckr}, but in such cases we should not expect strict dual resonance, due to the modified Regge behavior.}. 
For $q\geq 1$, this corresponds to a window $\left|1/(q\,{-}\,1)+t\right|<1/(q\,{-}\,1)$, within which the amplitude indeed obeys the unsubtracted dispersion relation in \Eq{eq:disp}.  In this case, the amplitude can be expressed in the dual resonant form in \Eq{amp_dual_res}, as advertised.
For $q<1$ the dual resonant sum in Eq.~\eqref{amp_dual_res} does not even converge in general, so we restrict to $q\geq 1$ throughout~\footnote{One could alternatively consider this object as an generalization of the Coon amplitude, and perhaps introduce contact terms to revive a less strict form of dual resonance.}.

Our solution~\eqref{eq:A} unifies many amplitudes previously discussed in the literature.  For  $\DD\,\,{=}\,\,0$, it is the $q$-hypergeometric amplitude of Ref.~\cite{Cheung:2023adk}, while further setting $r=0$ yields the Coon amplitude~\cite{Coon:1969yw}.
For $r=\DD=0$ and $q=1$, it becomes the Veneziano amplitude~\eqref{eq:AV}.

While $q\geq 1$ is permitted by crossing symmetry, we find that unitarity excludes $q>1$.  See \App{app:positivity} for a detailed analysis. Consequently, we henceforth consider $q=1$, corresponding to the linear string spectrum.  In this case, the amplitude in \Eq{eq:A} can be written as
\be
\hspace{-2.5mm}\begin{aligned}
A(s,t)&= \frac{1+r}{1\,{+}\,r\,{+}\,\DD}\frac{\Gamma(-s)\Gamma(-t)}{\Gamma(-s-t)}\biggl({}_{3}F_{2}\left[\begin{smallmatrix}
-s,\, -t,\, r\\
-s-t,\, 1+r
\end{smallmatrix} ;1\right]
\\& \;\;   +\frac{\DD(2\,{+}\,2r\,{+}\,\DD{+}\,s\,{+}\,t)}{(1+r)(s+t)}{}_{3}F_{2}\left[\begin{smallmatrix}
-s,\, -t,\, 1+r\\
1-s-t,\, 2+r
\end{smallmatrix} ;1\right]\biggr)
\\ &\; \; + \frac{w\left(2+2r+\DD\right)}{(1+r+\DD)st} ,
\\
\end{aligned}\hspace{-1mm} \label{eq:A1}
\ee
which is dual resonant for ${\rm Re}(t)<0$. 
\medskip

\mysubsec{High-Energy Behavior}Let us consider the behavior of the amplitude in \Eq{eq:A1} for high-energy scattering in the Regge limit, and for fixed angle, respectively.

In the Regge limit, the amplitude in \Eq{eq:A1} scales as
\eq{
A \sim s^{J(s,t)} + \frac{C(r,\DD)}{s(1+t)} + \frac{\DD}{s t(1+t)} + \cdots,
}{Regge_new}
where $C(r,\DD) = 1 + r + \DD - (1+r)/(1+r+\DD)$ and $J(s,t) =  t + \cdots$, with ellipses denoting subleading terms. 
For $t>-1$, the amplitude exhibits the $s^t$ scaling reminiscent of string theory, while for $t<-1$ it falls off as a power law.
The apparent $t$ channel pole is spurious, and simply indicates the transition between these two behaviors.

Crucially, we observe that superpolynomial softness for a range of $t$ holds if the terms in \Eq{Regge_new} vanish, which requires $r=\DD = 0$.
Hence, the unique amplitude that exhibits superpolynomial softness and level truncation is the Veneziano amplitude.

On the other hand, for hard fixed-angle scattering, where $|s|,|t| \rightarrow \infty$ at fixed $t/s$, \Eq{eq:A1} scales
\be 
A \sim e^{B(s,t)} +  \frac{C(r,\DD)}{s t} + \cdots, 
\ee 
where $B(s,t) =(s+t)\log(s+t)-s \log s - t \log t +\cdots $.   For $t>0$, the amplitude diverges as in the Veneziano case, while for $t<0$ there is power-law scaling.
\bigskip

\mysec{Discussion}We have presented a bootstrap problem that is analytically solvable and whose complete solution space spans the Veneziano amplitude and its cousins.  Our starting point is an arbitrary crossing symmetric amplitude with an unknown spectrum $\mm(n)$.  We then assume vanishing Regge behavior and level truncation, which mandates that at $t=\xx(k)$ the amplitude only receives contributions from levels $n<k$.  These conditions sculpt out a three-parameter $(q,r,w)$ space of solutions.  Only $q=1$ is consistent with positivity, and within this space only $r=w=0$ exhibits superpolynomial softness.  Thus, the unique object satisfying our most stringent conditions is the Veneziano amplitude.

Since our bootstrap conditions are very constraining, it is natural to relax our assumptions to accommodate more solutions.   
A natural deformation that  weakens our criteria is to instead assume  {\it partial} level truncation, which zeros out all exchanges at level $n\geq k+\lambda$, where $\lambda$ is a nonnegative integer and $\lambda=0$ corresponds to our original analysis.  Partial level truncation requires a more flexible form for the residue than \Eq{CRZ}.  For example, it is possible to consider the residue ansatz,
\eq{
R(n,t) = c(n) \prod_{k=1}^{n-\lambda} (t-\xx(k)) \prod_{j=n-\lambda+1}^{n} (t-\gamma(n,j)),
}{CRZ_lambda}
where $\gamma(n,j)$ are unknowns.  Plugging \Eq{CRZ_lambda} into \Eq{amp_dual_res} and enforcing the crossing symmetry conditions of \Eq{eq:crossing}, we obtain a system of nonlinear equations for the sequences of variables $c(n)$, $\xx(n)$, $\mm(n)$, and $\gamma(n,j)$.  

This system of equations is finite but not solvable using the straightforward analytic methods that were sufficient for the $\lambda=0$ case.  Indeed, the algebraic variety defined by these equations is sufficiently complicated that it approaches the limits of existing numerical techniques. 
It would be interesting to explore the varieties defined by our problem more systematically using the modern tools of algebraic geometry.

Nonetheless, since the Veneziano amplitude is a known solution of these nonlinear equations in the space of sequence variables, we can consider {\it linear perturbations} about it.  Expanding to linear order in perturbations of the sequence variables and finding the solutions that exhibit superpolynomial falloff---so $A(s,\xi(k)) \,{\sim}\, s^{-k}$ for all $k$---we find for $\lambda=0,1,2,3,4$ that there are $1, 5, 11, 18, 27$ such solutions, with spectra more general than the ($q$-deformed) integers.  While some or even all of these solutions may be spurious and could violate crossing symmetry at higher order in perturbations, it is enticing to imagine that these could be bona fide scattering amplitudes.  We leave a full analysis of these other putative solutions of the bootstrap to future work. 

Another possible direction for future work is the generalization to higher-point scattering.
As noted in \Eq{schematic_replacement}, the constraint of level truncation yields amplitudes that are structurally similar to the Veneziano amplitude but with kinematic poles and zeros shifted from integers to arbitrary sequences.  This observation immediately suggests a natural higher-point generalization.  For example, higher-point string amplitudes have well known integer-spaced poles, and their residues have numerous zeros and conform to stringent factorization constraints~\cite{Arkani-Hamed:2023swr, Arkani-Hamed:2023jwn, inprogress}.  It would be interesting to relocate these poles and zeroes to arbitrary locations to evaluate the uniqueness of higher-point string amplitudes.
Of course, the physical motivation for level truncation itself remains a compelling open question.

Last but not least is the question of extending our setup to include gravity.  Here the natural object of study is the Virasoro-Shapiro amplitude of string theory, which describes scattering gravitons.  Notably, the residues of the Virosoro-Shapiro amplitude exhibit {\it double zeroes}, in contrast with the single zeros that appear in \Eq{CRZ}.  For this reason it would be natural to explore whether gravitational amplitudes are uniquely specified by their Regge behavior and some version of enhanced level trunction.

\medskip
\noindent {\it Acknowledgments:} 
We thank Nima Arkani-Hamed, Justin Berman, Henriette Elvang, David Gross, Sebastian Mizera, Hirosi Ooguri, Simon Telen, and Sasha Zhiboedov for discussions. 
 C.C. and A.H. are supported by the Department of Energy (Grant No.~DE-SC0011632) and by the Walter Burke Institute for Theoretical Physics.
G.N.R. is supported by the James Arthur Postdoctoral Fellowship at New York University.

\onecolumngrid

\appendix

\section{Positivity}\label{app:positivity}

A necessary but not sufficient condition for unitarity is  positivity of the partial wave coefficients of the amplitude.
To study this condition, we perform a partial wave decomposition of the residue in \Eq{R_sol} in $D$ spacetime dimensions,
\be
R(n,t) = \sum_{\ell=0}^n a_{n,\ell} G_\ell^{(D)}(\cos\theta), 
\ee
where $G_\ell^{(D)}(\cos \theta)$ are Gegenbauer polynomials for scattering angle $\theta$ and 
\eq{
t = -\frac{1}{2} (s-4 \mu(0))(1-\cos\theta).
}{t_def}
Note that for our positivity analysis, we will reintroduce the parameter $\mu(0)$, which was fixed to zero earlier via the freedom of shifting the kinematic invariants $s$ and $t$.  To do so, we simply reverse this procedure, sending $s\rightarrow s-\mu(0)$ and $t\rightarrow t-\mu(0)$ in \Eq{R_sol} before expanding.  
Physically, $\mu(0)$  describes the zero point of the Regge trajectory, otherwise known as the mass squared of the lowest lying exchanged state.  We have chosen $\mu(0)$ to be the mass squared of the external states as well.

The partial wave coefficients for the general $(q,r,\DD)$ amplitude are then straightforwardly calculated. See the appendix of Ref.~\cite{Cheung:2022mkw} for various useful identities.
For $n>0$, we have
\be
\begin{aligned}
a_{n,\ell} & =\left(1+\frac{2\ell}{D-3}\right)\Gamma\left(\frac{D-1}{2}\right)\frac{q^n[1+n+r+\DD]_{q}}{[1+r+\DD]_{q}(q^{2+r};q)_{n}}\left[\sum_{k=\max(\ell,1)}^{n}\sum_{j=0}^{\lfloor\frac{k-\ell}{2}\rfloor}(\ell+2j)f(n,\ell,k,j)\right.\\
 & \;\;\qquad\qquad\qquad\qquad +\left.\sum_{k=\ell+1}^{n}\sum_{j=0}^{\lfloor\frac{k-\ell-1}{2}\rfloor}\frac{(k-\ell-2j)\left(2[-1-r-\DD]_{q}+[n]_{q}-\mu(0)\right)}{\frac{2}{1-q}+[n]_{q}-\mu(0)}f(n,\ell,k,j)\right]\\
f(n,\ell,k,j) & =\frac{[k]_{q}q^{k(k+2r+1)/2}(q;q)_{n}(q-1)^{k}}{[n]_{q}(q;q)_{k}(q;q)_{n-k}}\frac{(k-1)!}{(k-\ell-2j)!j!}\frac{\left(\frac{2}{1-q}+[n]_{q}-\mu(0)\right)^{k-\ell-2j}\left(3\mu(0)-[n]_{q}\right)^{\ell+2j}}{2^{k+\ell+2j}\Gamma\left(\frac{D-1}{2}+\ell+j\right)},
\end{aligned}
\ee
while for $n=0$, we have $a_{0,0} = 1$ in our normalization.
 In particular, for $n\geq 1$ the partial wave coefficients of the leading Regge trajectory at $\ell=n$ are
\be
a_{n,n}=\frac{\left(1+\tfrac{2n}{D-3}\right)n!(q-1)^{n}q^{n(n+3+2r)/2}}{\left(\tfrac{D-1}{2}\right)_{n}2^{2n}}\frac{[1+n+r+\DD]_{q}\left(3\mu(0)-[n]_{q}\right)^{n}}{[1+r+\DD]_{q}(q^{2+r};q)_{n}},
\ee
while those of the first subleading Regge trajectory at $\ell=n-1$ are given by
\be
a_{n,n-1}=\frac{(2n+d-5)\left[2n-(q-1)n\left([n]_{q}-\mu(0)\right)+2q^{-1-r}([1-n]_{q}-q^{-w})\right]}{n(q-1)\left([n]_{q}-3\mu(0)\right)} a_{n,n} .
\ee
For $q>1$, we see that as $n$ becomes large $a_{n,n-1} \sim -2na_{n,n}$, so as $n\rightarrow\infty$ it is impossible for $a_{n,n}$ and $a_{n,n-1}$ to have the same sign if $q>1$, for any values of $D$, $\mu(0)$, $r$, and $\DD$.
Since the dual resonant sum requires $q\geq 1$ in order to converge, unitarity---that is, the absence of ghosts---requires us to restrict to $q=1$.

\begin{figure*}[t]
\begin{center}
\includegraphics[width=0.4\columnwidth]{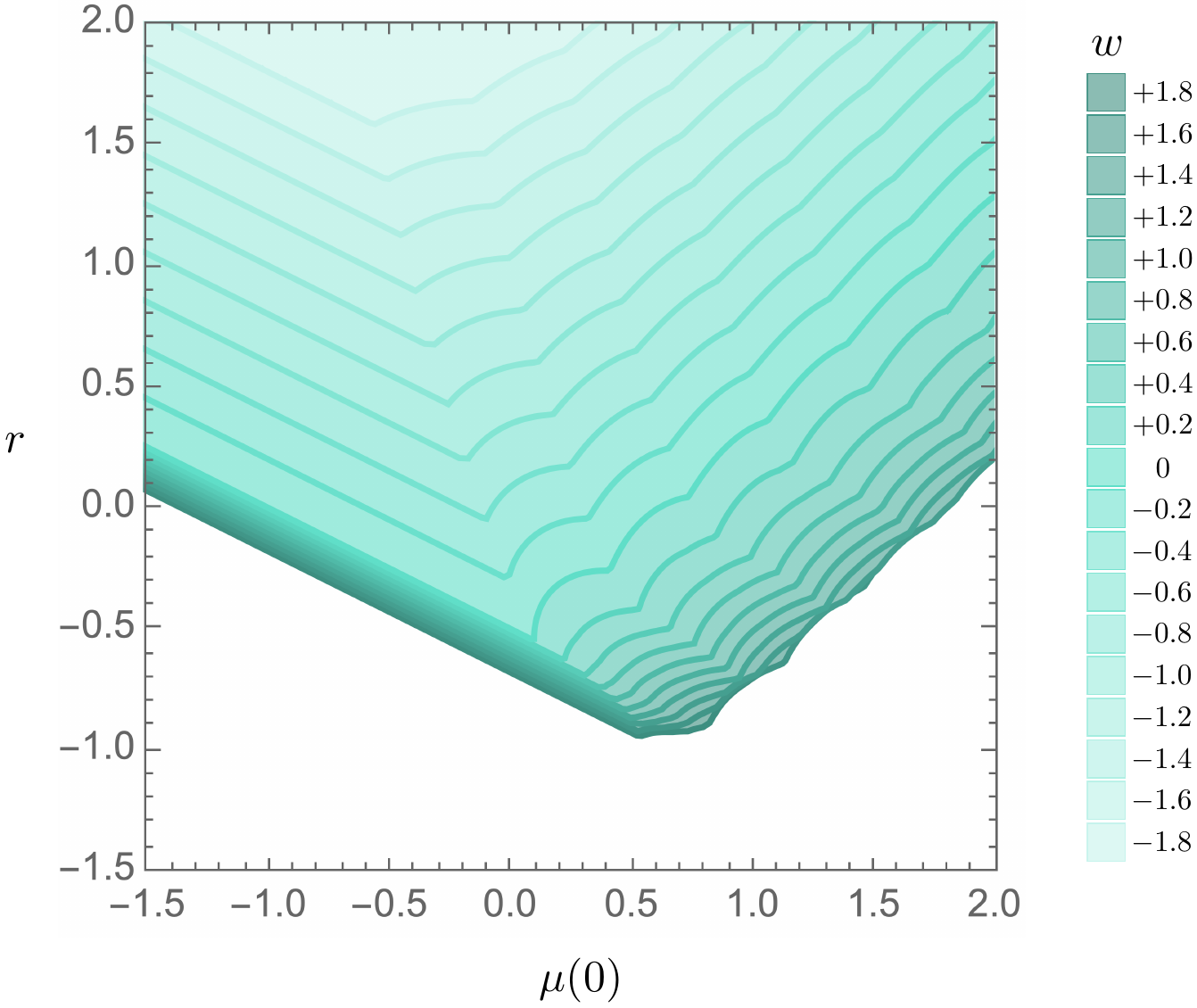}
\end{center}\vspace{-7mm}
\caption{Parameter space consistent with positivity for the amplitude in \Eq{eq:A1} with $q=1$ for various choices for $r$, $\DD$, and the mass squared of the external states, $\mu(0)$.  Here we have included positivity bounds up to maximum level $n=10$.  We have restricted to $D=4$ spacetime dimensions, but the positive region only diminishes for $D>4$, on account of the positivity of the connection coefficients of the Gegenbauer polynomials.
}
\label{fig:positivity}
\end{figure*}

Fixing $q$ to $1$, we find that the partial wave coefficients for $n>0$ are
\be 
\begin{aligned}a_{n,\ell} & =\left(1+\frac{2\ell}{D-3}\right)\Gamma\left(\frac{D-1}{2}\right)\frac{1+n+r+\DD}{(1+r+\DD)(2+r)_{n}}\left[\sum_{k=\max(\ell,1)}^{n}\sum_{j=0}^{\lfloor\frac{k-\ell}{2}\rfloor}(\ell+2j)f(n,\ell,k,j)\right.\\
 & \;\;\qquad\qquad\qquad\qquad\qquad+\left.\sum_{k=\ell+1}^{n}\sum_{j=0}^{\lfloor\frac{k-\ell-1}{2}\rfloor}\frac{(k-\ell-2j)(n-\mu(0)-2r-2-2\DD)}{n-\mu(0)-2r-4}f(n,\ell,k,j)\right]\\
f(n,\ell,k,j) & =(-1)^{k}S_{1}(n-1,k-1)\frac{(k-1)!}{(k-\ell-2j)!j!}\frac{\left(n-\mu(0)-2r-4\right)^{k-\ell-2j}\left(3\mu(0)-n\right)^{\ell+2j}}{2^{k+\ell+2j}\Gamma\left(\frac{D-1}{2}+\ell+j\right)},
\end{aligned}\label{eq:partials1}
\ee
where $S_1(n,k)$ is the unsigned Stirling number of the first kind.  As depicted in \Fig{fig:positivity},
there are broad regions of $(r,\DD,\mu(0))$ consistent with positivity.

Let us briefly comment on an important but underappreciated caveat with positivity analyses that is relevant to our specific setup.  Naively, partial wave unitarity requires that $a_{n,\ell} \geq 0$ for all levels $n$ and spins $\ell$. However, when $\mu(n) < 4\mu(0)$, the level $n$ state is below the physical threshold for production and can never go on shell.   Since this state is physically inaccessible, we should not place a positivity bound on its residue.  In particular, we should only enforce $a_{n,\ell} \geq 0$ for above-threshold states whose levels satisfy $\mu(n) \geq 4\mu(0)$~\cite{Arkani-Hamed:2023jwn}.  The crossover between physical and unphysical regimes at $\mu(n) = 4\mu(0)$ is encoded in a sign flip in \Eq{t_def} that in turn flips the sign of the corresponding residue for odd $\ell$.  So for example at level $n=1$, the naive positivity bound $\mu(0) \leq 1/3$ is a reflection of this crossover rather than of bona fide unitarity violation.  Said another way, partial wave unitarity for $n=1$ when $\mu(0) \leq 1/3$ is spurious because it applies to a below-threshold state. We emphasize that this subtlety is not a minor splitting of hairs.  Enforcing positivity for below-threshold states would rule out the standard model, since the $Z$ boson cannot be produced on shell from a pair of Higgs bosons due to the fact that $m_Z^2 < 4m_H^2$.

\twocolumngrid

\bibliographystyle{utphys-modified}
\bibliography{string_residue_zeros}

\providecommand{\href}[2]{#2}\begingroup\raggedright\begin{thebibliography}{10}

\bibitem{Elvang:2015rqa}
H.~Elvang and Y.-t. Huang, {\em {Scattering Amplitudes in Gauge Theory and
  Gravity}}.
\newblock Cambridge University Press, 2015.

\bibitem{Dixon:1996wi}
L.~J. Dixon, ``{Calculating scattering amplitudes efficiently},'' in {\em
  {Theoretical Advanced Study Institute in Elementary Particle Physics (TASI
  95): QCD and Beyond}}, pp.~539--584.
\newblock 1996.
\newblock \href{http://arxiv.org/abs/hep-ph/9601359}{{\ttfamily
  arXiv:hep-ph/9601359}}.

\bibitem{Cheung:2017pzi}
C.~Cheung, {\em {TASI Lectures on Scattering Amplitudes}},
  \href{http://dx.doi.org/10.1142/9789813233348_0008}{pp.~571--623}.
\newblock 2018.
\newblock \href{http://arxiv.org/abs/1708.03872}{{\ttfamily arXiv:1708.03872
  [hep-ph]}}.

\bibitem{Travaglini:2022uwo}
G.~Travaglini { et~al.}, ``{The SAGEX review on scattering amplitudes},''
  \href{http://dx.doi.org/10.1088/1751-8121/ac8380}{{\em J. Phys. A} {\bfseries
  55} (2022) 443001}, \href{http://arxiv.org/abs/2203.13011}{{\ttfamily
  arXiv:2203.13011 [hep-th]}}.

\bibitem{Note1}
Throughout this paper we consider planar, color-ordered amplitudes, which arise
  in gauge theories or open string theory.

\bibitem{Cheung:2023adk}
C.~Cheung and G.~N. Remmen, ``{Stringy dynamics from an amplitudes
  bootstrap},'' \href{http://dx.doi.org/10.1103/PhysRevD.108.026011}{{\em Phys.
  Rev. D} {\bfseries 108} (2023) 026011},
  \href{http://arxiv.org/abs/2302.12263}{{\ttfamily arXiv:2302.12263
  [hep-th]}}.

\bibitem{Caron-Huot:2016icg}
S.~Caron-Huot, Z.~Komargodski, A.~Sever, and A.~Zhiboedov, ``{Strings from
  Massive Higher Spins: The Asymptotic Uniqueness of the Veneziano
  Amplitude},'' \href{http://dx.doi.org/10.1007/JHEP10(2017)026}{{\em JHEP}
  {\bfseries 10} (2017) 026}, \href{http://arxiv.org/abs/1607.04253}{{\ttfamily
  arXiv:1607.04253 [hep-th]}}.

\bibitem{Cheung:2023uwn}
C.~Cheung and G.~N. Remmen, ``{Bespoke dual resonance},''
  \href{http://dx.doi.org/10.1103/PhysRevD.108.086009}{{\em Phys. Rev. D}
  {\bfseries 108} (2023) 086009},
  \href{http://arxiv.org/abs/2308.03833}{{\ttfamily arXiv:2308.03833
  [hep-th]}}.

\bibitem{Nakanishi:1970xz}
N.~Nakanishi, ``{Four-Point and Five-Point Veneziano-Type Formulas on the Basis
  of Spectral Representations},''
  \href{http://dx.doi.org/10.1103/PhysRevD.2.288}{{\em Phys. Rev. D} {\bfseries
  2} (1970) 288}.

\bibitem{Matsuda:1969cso}
S.~Matsuda, ``{Uniqueness of the Veneziano Representation},''
  \href{http://dx.doi.org/10.1103/PhysRev.185.1811}{{\em Phys. Rev.} {\bfseries
  185} (1969) 1811}.

\bibitem{Huang:2022mdb}
Y.-t. Huang and G.~N. Remmen, ``{UV-complete gravity amplitudes and the triple
  product},'' \href{http://dx.doi.org/10.1103/PhysRevD.106.L021902}{{\em Phys.
  Rev. D} {\bfseries 106} (2022) L021902},
  \href{http://arxiv.org/abs/2203.00696}{{\ttfamily arXiv:2203.00696
  [hep-th]}}.

\bibitem{Bern:2021ppb}
Z.~Bern, D.~Kosmopoulos, and A.~Zhiboedov, ``{Gravitational effective field
  theory islands, low-spin dominance, and the four-graviton amplitude},''
  \href{http://dx.doi.org/10.1088/1751-8121/ac0e51}{{\em J. Phys. A} {\bfseries
  54} (2021) 344002}, \href{http://arxiv.org/abs/2103.12728}{{\ttfamily
  arXiv:2103.12728 [hep-th]}}.

\bibitem{Caron-Huot:2020cmc}
S.~Caron-Huot and V.~Van~Duong, ``{Extremal Effective Field Theories},''
  \href{http://dx.doi.org/10.1007/JHEP05(2021)280}{{\em JHEP} {\bfseries 05}
  (2021) 280}, \href{http://arxiv.org/abs/2011.02957}{{\ttfamily
  arXiv:2011.02957 [hep-th]}}.

\bibitem{Caron-Huot:2021rmr}
S.~Caron-Huot, D.~Mazac, L.~Rastelli, and D.~Simmons-Duffin, ``{Sharp
  boundaries for the swampland},''
  \href{http://dx.doi.org/10.1007/JHEP07(2021)110}{{\em JHEP} {\bfseries 07}
  (2021) 110}, \href{http://arxiv.org/abs/2102.08951}{{\ttfamily
  arXiv:2102.08951 [hep-th]}}.

\bibitem{Martin}
F.~A. Cerulus and A.~Martin, ``{A lower bound for large angle elastic
  scattering at high energies},''
  \href{http://dx.doi.org/10.1016/0031-9163(64)90807-8}{{\em Phys. Lett.}
  {\bfseries 8} (1964) 80}.

\bibitem{Buoninfante:2023dyd}
L.~Buoninfante, J.~Tokuda, and M.~Yamaguchi, ``{New lower bounds on scattering
  amplitudes: non-locality constraints},''
  \href{http://dx.doi.org/10.1007/JHEP01(2024)082}{{\em JHEP} {\bfseries 01}
  (2024) 082}, \href{http://arxiv.org/abs/2305.16422}{{\ttfamily
  arXiv:2305.16422 [hep-th]}}.

\bibitem{Haring:2023zwu}
K.~H\"aring and A.~Zhiboedov, ``{The Stringy S-matrix Bootstrap: Maximal Spin
  and Superpolynomial Softness},''
  \href{http://arxiv.org/abs/2311.13631}{{\ttfamily arXiv:2311.13631
  [hep-th]}}.

\bibitem{Note2}
If the sequence $t_k$ is unboundedly negative for large $k$ then $t=t_k$
  interpolates to the regime of high-energy fixed-angle scattering. In this
  case, superpolynomial softness coincides with exponentially soft fixed-angle
  scattering, which is another calling card of string amplitudes. A loophole to
  this logic is if $t_k$ approaches an accumulation point, which actually
  occurs for the Coon amplitude when $q>1$.

\bibitem{Note3}
The Veneziano amplitude referred to here and throughout describes the
  scattering of external states that are massless colored scalars. This
  amplitude is obtained by a trivial affine shift of $s$ and $t$ in the bosonic
  string amplitude or, alternatively, by stripping off the polarization data
  from superstring scattering, which is also known as the $Z$-theory
  amplitude~\cite {Carrasco:2016ldy}.

\bibitem{Coon:1969yw}
D.~D. Coon, ``{Uniqueness of the Veneziano representation},''
  \href{http://dx.doi.org/10.1016/0370-2693(69)90106-3}{{\em Phys. Lett. B}
  {\bfseries 29} (1969) 669}.

\bibitem{Gross:1969db}
D.~J. Gross, ``{Factorization and the generalized Veneziano model with
  satellites},'' \href{http://dx.doi.org/10.1016/0550-3213(69)90248-X}{{\em
  Nucl. Phys. B} {\bfseries 13} (1969) 467}.

\bibitem{Eckner:2024ggx}
C.~Eckner, F.~Figueroa, and P.~Tourkine, ``{The Regge bootstrap, from linear to
  non-linear trajectories},'' \href{http://arxiv.org/abs/2401.08736}{{\ttfamily
  arXiv:2401.08736 [hep-th]}}.

\bibitem{Note4}
The $q$-deformed (basic) generalized hypergeometric functions are defined by
  the sum \begin {equation*} \protect \tmspace +\thickmuskip {.2777em}\hskip
  2em\relax {}\protect \,_{n+1}\phi _{n}\left [\begin {smallmatrix}
  a_1,\protect \ldots ,a_{n+1}\\ b_1,\protect \ldots ,b_n \end
  {smallmatrix};q;z\right ] = {\scalebox {1.85}{\protect \raisebox
  {-0.65mm}{$\Sigma $}}}_{k=0}^\infty \protect \genfrac {}{}{}1{(a_1;q)_k
  \protect \cdots (a_{n+1};q)_k}{(b_1;q)_k \protect \cdots (b_n;q)_k}\protect
  \genfrac {}{}{}1{z^k}{(q;q)_k}, \end {equation*} and $\Gamma
  _q(z)=(1-q)^{1-z}(q;q)_\infty /(q^z;q)_\infty $ is the $q$-gamma function.

\bibitem{Note5}
A remarkable feature of this amplitude is the appearance of the $q^{\sigma \tau
  }$ prefactor. This very same quantity was introduced as a nonunique~\cite
  {Jepsen:2023sia} fudge factor sitting in front of the amplitude in Ref.~\cite
  {Coon:1972qz} in order to cancel the nonpolynomiality of the residues of the
  original Coon amplitude~\cite {Coon:1969yw} for $q<1$. Surprisingly, we find
  that this same prefactor arises {\protect \it automatically} from the
  meromorphic dual resonant sum for the amplitude for $q>1$, and need not be
  included by hand. Meanwhile, in the high-energy fixed-angle limit defined by
  $|s|,|t|\rightarrow \infty $ at fixed $t/s$, we find that $\protect \qopname
  \relax o{log}A \sim \protect \qopname \relax o{log}s \protect \qopname \relax
  o{log}t/\protect \qopname \relax o{log}q$ for $q > 1$. This is the
  distinctive double-log scaling characteristic of the Coon amplitude~\cite
  {Caron-Huot:2016icg,Cheung:2022mkw}, but which we have found here to be
  independent of the new parameters $r$ and $w$. Notably, such behavior for
  $q<1$ can arise in consistent ultraviolet theories, such as the anti-de
  Sitter/D-brane construction of Ref.~\cite {Maldacena:2022ckr}, but in such
  cases we should not expect strict dual resonance, due to the modified Regge
  behavior.

\bibitem{Note6}
One could alternatively consider this object as an generalization of the Coon
  amplitude, and perhaps introduce contact terms to revive a less strict form
  of dual resonance.

\bibitem{Arkani-Hamed:2023swr}
N.~Arkani-Hamed, Q.~Cao, J.~Dong, C.~Figueiredo, and S.~He, ``{Hidden zeros for
  particle/string amplitudes and the unity of colored scalars, pions and
  gluons},'' \href{http://arxiv.org/abs/2312.16282}{{\ttfamily arXiv:2312.16282
  [hep-th]}}.

\bibitem{Arkani-Hamed:2023jwn}
N.~Arkani-Hamed, C.~Cheung, C.~Figueiredo, and G.~N. Remmen, ``{Multiparticle
  Factorization and the Rigidity of String Theory},''
  \href{http://dx.doi.org/10.1103/PhysRevLett.132.091601}{{\em Phys. Rev.
  Lett.} {\bfseries 132} (2024) 091601},
  \href{http://arxiv.org/abs/2312.07652}{{\ttfamily arXiv:2312.07652
  [hep-th]}}.

\bibitem{inprogress}
N.~Arkani-Hamed, C.~Cheung, C.~Figueiredo, and G.~N. Remmen. In progress.

\bibitem{Cheung:2022mkw}
C.~Cheung and G.~N. Remmen, ``{Veneziano variations: how unique are string
  amplitudes?},'' \href{http://dx.doi.org/10.1007/JHEP01(2023)122}{{\em JHEP}
  {\bfseries 01} (2023) 122}, \href{http://arxiv.org/abs/2210.12163}{{\ttfamily
  arXiv:2210.12163 [hep-th]}}.

\bibitem{Carrasco:2016ldy}
J.~J.~M. Carrasco, C.~R. Mafra, and O.~Schlotterer, ``{Abelian $Z$-theory: NLSM
  amplitudes and $\alpha'$-corrections from the open string},''
  \href{http://dx.doi.org/10.1007/JHEP06(2017)093}{{\em JHEP} {\bfseries 06}
  (2017) 093}, \href{http://arxiv.org/abs/1608.02569}{{\ttfamily
  arXiv:1608.02569 [hep-th]}}.

\bibitem{Jepsen:2023sia}
C.~B. Jepsen, ``{Cutting the Coon amplitude},''
  \href{http://dx.doi.org/10.1007/JHEP06(2023)114}{{\em JHEP} {\bfseries 06}
  (2023) 114}, \href{http://arxiv.org/abs/2303.02149}{{\ttfamily
  arXiv:2303.02149 [hep-th]}}.

\bibitem{Coon:1972qz}
D.~D. Coon, U.~P. Sukhatme, and J.~Tran Thanh~Van, ``{Duality and proton-proton
  scattering at all angles},''
  \href{http://dx.doi.org/10.1016/0370-2693(73)90205-0}{{\em Phys. Lett. B}
  {\bfseries 45} (1973) 287}.

\bibitem{Maldacena:2022ckr}
J.~Maldacena and G.~N. Remmen, ``{Accumulation-point amplitudes in string
  theory},'' \href{http://dx.doi.org/10.1007/JHEP08(2022)152}{{\em JHEP}
  {\bfseries 08} (2022) 152}, \href{http://arxiv.org/abs/2207.06426}{{\ttfamily
  arXiv:2207.06426 [hep-th]}}.

\end{thebibliography}\endgroup

\end{document}